# Low temperature behavior of the heavy Fermion $Ce_3Co_4Sn_{13}$

A. D. Christianson[a,b,c], J. S. Gardner[d,e], H. J. Kang[d,f], J.-H. Chung[d,f], S. Bobev[g], J. L. Sarrao[c], and J. M. Lawrence[b]

[a]Oak Ridge National Laboratory, Oak Ridge, Tennessee 37831, USA
[b]University of California, Irvine, California 92697, USA
[c]Los Alamos National Laboratory, Los Alamos, New Mexico 87545, USA
[d]NIST Center for Neutron Research, Gaithersburg, Maryland 20899, USA
[e]IUCF, Indiana University, Bloomington, Indiana 47405, USA
[f]University of Maryland, College Park, Maryland 20742, USA
[g]University of Delaware, Newark, Deleware 19716, USA

The compound $Ce_3Co_4Sn_{13}$ is an extremely heavy cubic heavy fermion system with a low temperature electronic specific heat of order ~4 J/mol-$K^2$. If the compound is nonmagnetic, it would be one of the heaviest nonmagnetic Ce-based heavy fermions reported to date and therefore would be expected to lie extremely close to a quantum critical point. However, a broad peak of unknown origin is observed at 0.8 K in the specific heat and magnetic susceptibility, suggesting the possibility of antiferromagnetic order. We present neutron diffraction data from polycrystalline samples which do not show any sign of magnetic scattering below 0.8 K. In addition, we present inelastic neutron scattering data from a single crystal sample which is consistent with the 1.2 K energy scale for Kondo spin fluctuations determined from specific heat measurements.

The behavior of the antiferromagnetic quantum critical point (QCP) in heavy fermion systems remains a controversial topic in condensed matter physics. In $CeRu_2Si_2$ inelastic neutron scattering investigations point to a spin density wave type quantum critical point.[1,2,3] In $CeCu_6$, E/T scaling is observed for the antiferromagnetic fluctuations suggesting a local QCP.[4] To identify the universal aspects of the heavy Fermi QCP, further examples are needed, ideally systems of high symmetry and systems which do not require chemical substitution and the associated effects of disorder.

$Ce_3Co_4Sn_{13}$ is a new cubic heavy fermion material with a low temperature electronic specific heat as high as 4 J/mol $K^2$.[5,6] Specific heat measurements show a broad peak at 0.8 K.[6] This feature is broader than normally expected for a long range antiferromagnetic transition as well as that for a Schottky anomaly. The possibility that the feature at 0.8 K is due to low lying crystal field levels is further reduced by the following argument: the $Ce^{3+}$ site possesses tetragonal symmetry as opposed to the overall cubic symmetry of $Ce_3Co_4Sn_{13}$ and thus the 6-fold degeneracy of the ground state multiplet is lifted under the influence of the crystalline environment to that of 3 doublets. Moreover, inelastic neutron scattering experiments show ground state crystal field

excitations at 8 and 30 meV accounting for all of the crystal field levels.[7]

Specific heat measurements under applied magnetic fields as high as 9 T show the peak broadens and moves upward in temperature.[6] Fitting the data to the S = ½ Kondo impurity model yields a Kondo temperature ($T_K$) of 1.2 K. To examine the low temperature behavior of $Ce_3Co_4Sn_{13}$ in greater detail, we have performed neutron diffraction measurements to search for magnetic order as well as inelastic neutron scattering experiments to verify the Kondo energy scale.

Single crystals of $Ce_3Co_4Sn_{13}$ were grown in Sn-flux as described elsewhere[5]. Samples from batches prepared under identical conditions to those used here were characterized by single crystal x-ray diffraction at room temperature and magnetic susceptibility measurements. Neutron powder diffraction data were collected for 8 hours at 0.4 and 2 K in a $^3$He system on a 10 g sample of crushed single crystals using the BT-1 powder diffractometer (Cu(311) monochromator, $\lambda$ = 1.5403(2) Å) at the NIST Center for Neutron Research (NCNR). Data were collected over the 2$\Theta$ range of 3-168°. Inelastic neutron scattering experiments were performed on a 4.8 g single crystal in a pumped He cryostat on the cold neutron spectrometer SPINS at the NCNR. The pyrolytic graphite (002) reflection was used for monochromator and analyzer. The collimations were 40'-S-40'-open and the final energy was fixed at $E_f$ = 3.0 meV yielding an energy resolution of 0.08 meV (FWHM). A Be filter was placed after the sample to remove higher order energy contamination. Several energy scans were made and summed giving a total counting time of 45 minutes per point.

Fig. 1 shows the result of subtracting the 2 K diffraction pattern from the 0.4 K diffraction pattern at low scattering angles. Within the error bars no extra scattering is observed which can be attributed to magnetic ordering. Although an estimate of the maximum moment is model dependent if the moment were larger than 0.8 $\mu_B$ a noticeable difference should be evident in fig. 1. There are several aspects of the physical behavior of $Ce_3Co_4Sn_{13}$ which may preclude the observation of long range magnetic order. The most important is that for a heavy Fermion material with a large low temperature specific heat as in $Ce_3Co_4Sn_{13}$ the magnetic moments are expected to be small and hence may be unobservable in a powder diffraction experiment. Another possibility is that if the specific heat peak were due in part to short range magnetic order any elastic magnetic response would be broadened and difficult to observe. Further investigations with single crystals are underway to clarify this point.

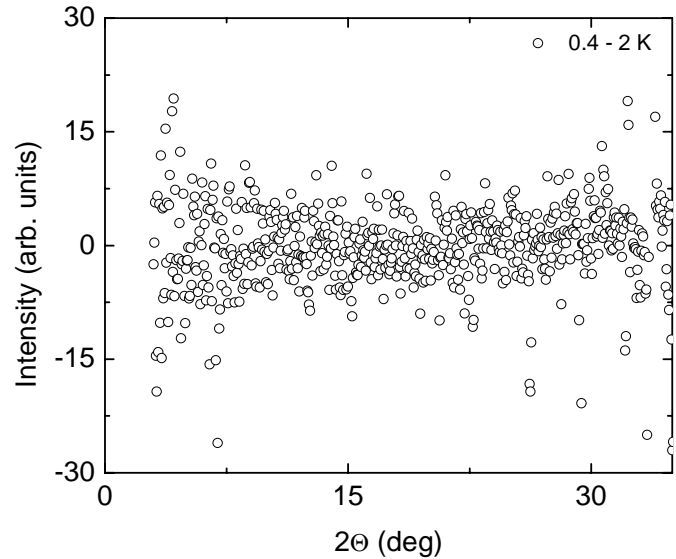

Fig. 1. The result of subtracting neutron diffraction data collected at 2 K from data collected at 0.4 K. The error bars are approximately indicated by the scatter of the data and have been omitted for clarity.

We now turn to the results of inelastic neutron scattering experiments on a single crystal. Fig. 2 shows an energy scan at $\mathbf{q} = (0.5\ 0.5\ 0.5)$ with the crystal orientated with the [1 -1 0] direction vertical. The line at the bottom is a measurement of the empty holder with experimental conditions as described above. The peak at the elastic line is due to the incoherent scattering of the sample. The inset shows the result of subtracting the empty holder as well as fits to the data. The result of subtracting the empty holder from the sample scattering can be fit assuming a Lorentzian line shape that is either quasielastic or inelastic (fig. 2). Although, the statistics and resolution of this experiment prevent differentiation between either of these two cases, in each case the energy scale (1.4 K for quasielastic fits and 2.6 K for inelastic fits) is in reasonable agreement with the $T_K = 1.2$ K derived from the specific heat measurements of ref. [6]. Further experiments with a higher resolution and longer counting times will be required to draw more detailed conclusions about the Kondo spin fluctuations in $Ce_3Co_4Sn_{13}$.

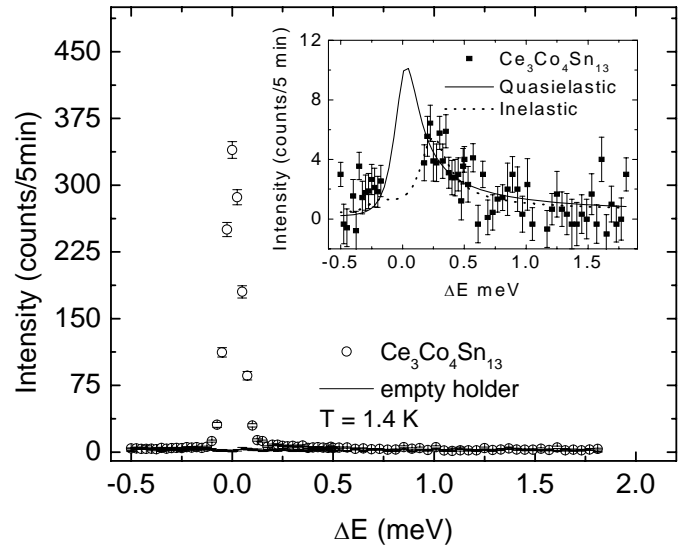

Fig. 2. Energy scan at **q** = (0.5 0.5 0.5) at 1.4 K. The inset shows the result of subtracting the empty holder. The solid (dotted) line is a quasielastic (inelastic) fit.

Work at UC Irvine was supported by the Department of Energy (DOE) under Grant No. DE-FG03-03ER46036. Oak Ridge National Laboratory is managed by UT-Battelle, for the DOE under Contract No. DE-AC05-00OR22725. Work at Los Alamos was performed under the auspices of the DOE. This work utilized facilities supported in part by the NSF under Agreement No. DMR-0454672

**References**
[1] L.P. Regnault, *et al*., Phys. Rev. B **38**, 4481 (1988).
[2] S. Raymond, *et al*., *Prog. Theor. Phys.* **32**, 37 (1964).
[3] H. Kadowaki, *et al*., Phys. Rev. Lett. **96**, 016401 (2006).
[4] A. Schröder. *et al*., Nature **407**, 351 (2000).
[5] C. Israel, *et al*., Physica B **359-361**, 251 (2005).
[6]A. L. Cornelius, *et al*., Physica B **378-380**, 113 (2006).
[7]A. D. Christianson, *et al*., unpublished.